\documentclass{article}
\usepackage{spconf,amsmath,amssymb,graphicx,hyperref}

\title{VoTSpeech: Decoupling Voice Design from Speech Generation in Instruction-Following TTS}
\name{
    Wenbing Yang$^{1}$,
    Qihang Lu$^{1}$,
    Zihan Sun$^{2}$,
    Peilei Jia$^{2}$,
    Yingming Gao$^{1}$,
    Ya Li$^{1,*}$,
    Jun Gao$^{2,*}$
    \thanks{*Corresponding author.}
}

\address{
    $^{1}$Beijing University of Posts and Telecommunications, Beijing, China \\
    $^{2}$Hello Group, Beijing, China
}
\begin{document}
%
\maketitle
\begin{abstract}
Instruction-following text-to-speech (TTS) typically leaves voice design implicit, as conventional speaker embeddings encode reference audio rather than text-driven targets. We propose VoTSpeech, which decouples voice design from speech generation via an explicit intermediate voice latent. Built on a shared language model backbone, VoTSpeech first predicts a continuous voice latent from text instructions using a flow-based module. This latent serves as a shared condition for both the language model and the Acoustic DiT, guiding autoregressive sequence modeling and acoustic latent generation, respectively, before an audio VAE decoder converts the acoustic latents into waveforms. To capture speaker identity and complementary acoustic nuances, the voice latent is jointly supervised by speaker embeddings and temporally pooled audio VAE features. With task-specific adaptation, VoTSpeech performs strongly on InstructTTSEval-ZH, demonstrating competitive capabilities against strong open-source models while maintaining high naturalness. These results confirm that explicit voice latent design and its conditioning interface effectively empower text-driven voice design for controllable speech synthesis. Code, model, and audio samples are available at \url{https://github.com/ywbn/VoTSpeech}.
\end{abstract}

\begin{keywords}
instruction-following TTS, voice design, decoupled voice modeling, flow matching.
\end{keywords}

\section{Introduction}
\label{sec:intro}

Natural-language voice design enables personalized speech synthesis directly from textual descriptions without reference audio \cite{prompttts, instructtts, voicesculptor}. Despite recent advances in instruction-following TTS \cite{mingomni, qwen3tts, moss, voxcpm2}, most existing models simply feed voice instructions as textual context directly into TTS systems. Without an explicit voice latent, target voice creation remains implicit within utterance generation. This coupling obscures explicit control, motivating a principled separation between \textit{what voice to create} and \textit{what that voice says}.

An explicit continuous voice latent bridges text prompts and speech generation. While reference-based TTS reliably extracts speaker representations from audio \cite{jia2018transfer, du2024cosyvoice, zhou2026indextts2}, text-driven synthesis must construct target representations without audio references. Prior studies have explored generative speaker embeddings \cite{stanton2022speaker, promptspeaker} or description-conditioned voice representations \cite{leng2024prompttts, MiniMaxSpeech, captalk}. Yet, many existing representations focus primarily on static speaker identity, neglecting dynamic acoustic nuances crucial for expressive synthesis. Consequently, an effective intermediate latent must balance semantic description alignment with fine-grained acoustic richness. Predicting such a multi-faceted voice latent and integrating it effectively into speech generation within a unified architecture remains a non-trivial challenge.

To address this challenge, we propose \textbf{VoTSpeech} (Voice-of-Thought Speech), which decouples voice design from speech generation via an explicit continuous voice latent. Adopting a ``think-before-speaking'' paradigm, a shared large language model (LLM) first interprets instructions to predict a target voice latent via a flow-based module. This latent then conditions both the LLM and the Acoustic Diffusion Transformer (DiT), guiding sequence prosody planning and acoustic rendering, respectively. For balanced supervision, the target latent is jointly trained on speaker embeddings and temporally pooled audio Variational Autoencoder (VAE) features. Evaluations on InstructTTSEval-ZH \cite{instructttseval} show that VoTSpeech achieves highly competitive performance against strong open-source baselines, notably excelling in descriptive-style and role-play tasks.
Our contributions are summarized as follows:
\begin{itemize}
    \item We propose VoTSpeech, a framework decoupling instruction-guided voice design from speech generation via an explicit continuous voice latent.
    \item We design a dual-path conditioning interface supervised by a hybrid target combining speaker embeddings and pooled audio VAE features to empower both LLM prosody planning and DiT acoustic rendering.
    \item We develop a consistency-controlled annotation pipeline and demonstrate VoTSpeech's competitive performance on InstructTTSEval-ZH, with ablations validating the core design choices.
\end{itemize}

\begin{figure*}[t]
    \centering
    \includegraphics[width=0.8\textwidth]{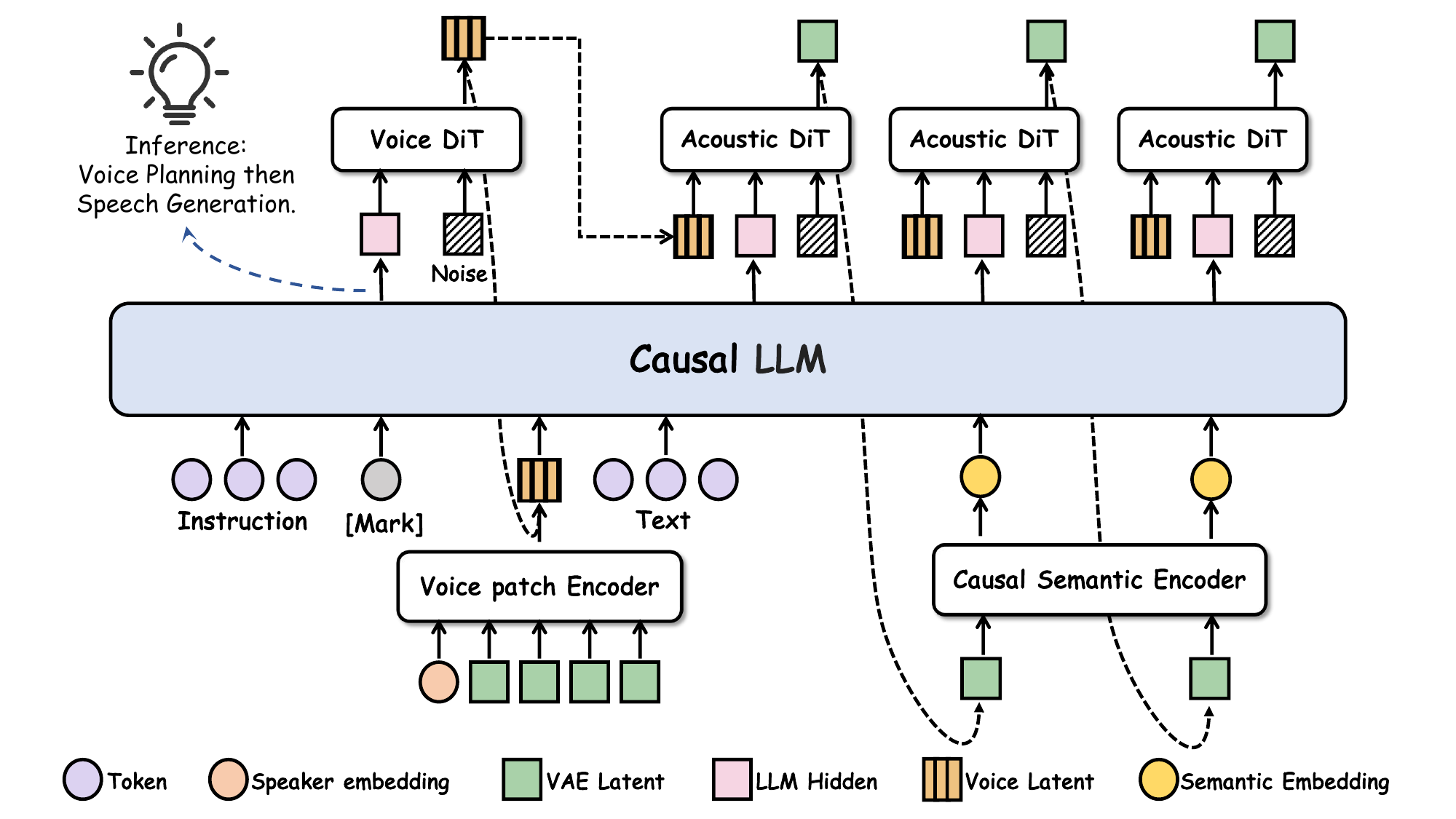}
    \caption{Overview of VoTSpeech. Voice DiT predicts a continuous voice latent from an instruction-derived LLM state. The resulting latent conditions both the LLM and Acoustic DiT through dual pathways and is trained against a speech-derived target combining speaker embeddings and pooled audio VAE features.}
    \label{fig:architecture}
\end{figure*}

\section{Method}
\label{sec:method}

VoTSpeech decouples voice design from speech generation via an explicit continuous voice latent. Supported by an automated, quality-controlled instruction annotation pipeline, a shared Causal LLM interprets instructions to predict the voice latent via a Voice DiT, which subsequently conditions both LLM prosody planning and Acoustic DiT rendering (Fig.~\ref{fig:architecture}). Building upon dots.tts \cite{dotstts}, VoTSpeech introduces a Voice Patch Encoder and Voice DiT for explicit voice modeling. 

\subsection{Instruction Data Construction}
\label{ssec:data}

To distinguish persistent voice identity from dynamic speech delivery, we construct instruction--speech pairs across two complementary annotation levels:

\textbf{Speaker-Level Inherent Trait Annotation:} Multiple recordings from the same speaker are grouped and processed together to capture persistent attributes (e.g., perceived age, gender, pitch register, vocal texture, and accent), following ParaSpeechCaps \cite{paraspeechcaps}. The resulting verified description is assigned to all utterances belonging to that speaker.

\textbf{Utterance-Level Dynamic Style Annotation:} Individual recordings are annotated as free-form, actionable control instructions detailing localized delivery dynamics. Rather than relying on rigid templates, these descriptions highlight salient acoustic attributes, including pitch range, stress and volume placement, tempo progression, and emotional trajectory.

PromptDialog \cite{promptdialog_dataset} and Genshin \cite{genshin_dataset} recordings provide source data for these annotations, complemented by rewritten tag descriptions from ISCSLP2026-CoT-TTS \cite{iscslpcottts}. To improve annotation quality and reduce semantic drift in LLM-generated text \cite{geng2026stabilizing}, Qwen3-Omni \cite{qwen3omni} directly processes the audio inputs (multi-utterance groups for speaker-level and single recordings for utterance-level) to sample two candidate descriptions per instance. DeepSeek-V4 \cite{deepseekv4} then evaluates attribute-level consistency between candidates: instances with major contradictions are discarded, whereas localized conflicts are resolved by pruning ambiguous attributes and retaining consensus content.

\subsection{Model Architecture}
\label{ssec:architecture}

\textbf{Sequence organization.}
VoTSpeech introduces a voice-design stage before synthesis text processing, with the sequence organized as:
\begin{equation}
    [\,c,\ \langle v\rangle,\ e_v,\ x,\ \ell_{1:T}\,],
    \label{eq:sequence}
\end{equation}
where $c$ denotes the voice instruction, $\langle v\rangle$ the special voice-generation marker, $e_v$ the continuous voice embedding, $x$ the synthesis text, and $\ell_{1:T}$ the speech-latent sequence, represented in the LLM by Causal Semantic Encoder outputs. Under causal attention, the hidden state $h_c$ at $\langle v\rangle$ summarizes the instruction prefix without accessing target text $x$ and conditions the flow-based voice latent prediction. Inserting $e_v$ before $x$ allows text processing and subsequent speech generation to attend to the designed voice, while the instruction context remains accessible for expressive control.

\textbf{Voice Patch Encoder.}
While reference-conditioned frameworks \cite{mingomni, qwen3tts, dotstts} rely solely on a speaker embedding $E_{\mathrm{spk}}(y)$ to represent voice identity, text-driven voice design requires capturing both speaker identity and global acoustic nuances without reference audio. To construct a multi-faceted target latent $z^*$ without extra disentanglement objectives \cite{qian2019autovc, lu2023disentangled}, the Voice Patch Encoder augments speaker identity with temporally pooled acoustic features from waveform $y$:
\begin{equation}
    z^* = \operatorname{norm}\!\left(
    \left[E_{\mathrm{spk}}(y);
    \operatorname{pool}\!\left(E_{\mathrm{vae}}(y)\right)\right]
    \right),
    \label{eq:voice_code}
\end{equation}
where $E_{\mathrm{spk}}$ is the CAM++ encoder \cite{CAM++}, $E_{\mathrm{vae}}$ is the audio VAE encoder, $\operatorname{pool}$ denotes temporal mean-pooling over valid frames, and $\operatorname{norm}$ applies per-dimension standardization using dataset statistics. Conditioned on instruction context $h_c$, the flow-matching Voice DiT predicts continuous voice latent $z$ from Gaussian noise using $z^*$ as its training target.

\textbf{Dual-path Voice Conditioning.}
The original architecture injects voice representations solely into the acoustic decoder, leaving language model prosody planning without explicit voice guidance. VoTSpeech overcomes this limitation by leveraging the instruction-predicted latent $z$ to establish a dual-path conditioning interface, guiding both LLM sequence modeling and Acoustic DiT rendering simultaneously:
\begin{equation}
    \begin{aligned}
        e_v &= P_z(z) + P_h(h_c), \\
        g_v &= G_s(\tilde{s}) + G_a(\tilde{a}) + G_h(h_c),
    \end{aligned}
    \label{eq:conditioning}
\end{equation}
where $[\tilde{s}; \tilde{a}] = \operatorname{norm}^{-1}(z)$ recovers speaker and acoustic components, and $P_z, P_h, G_s, G_a, G_h$ are learned linear projections. The prefix embedding $e_v$ prepends synthesis text $x$ in the LLM, while $g_v$ conditions the Acoustic DiT, ensuring unified voice guidance across sequence modeling and acoustic rendering. Zero-initialized residuals $P_h$ and $G_h$ provide direct instruction context for additional stability.

\subsection{Training and Inference}
\label{ssec:training}

\textbf{Joint Optimization and Gradient Decoupling.} We train Voice DiT and the speech generation model jointly while keeping encoders $E_{\mathrm{spk}}$ and $E_{\mathrm{vae}}$ frozen. Voice DiT learns a conditional vector field mapping Gaussian noise $\epsilon \sim \mathcal{N}(0, \mathbf{I})$ to the target voice latent $z^*$. For time step $t \sim \mathcal{U}(0,1)$ and interpolated state $z_t = (1-t)\epsilon + t z^*$, the voice prior objective is defined as:
\begin{equation}
    \mathcal{L}_{\mathrm{voice}} = \mathbb{E}\!\left[ \left\| v_{\theta}\!\left(z_t, t, \operatorname{sg}(h_c)\right) - (z^* - \epsilon) \right\|_w^2 \right],
    \label{eq:voice_loss}
\end{equation}
where $\operatorname{sg}(\cdot)$ denotes the stop-gradient operator. Crucially, $\operatorname{sg}(h_c)$ prevents $\mathcal{L}_{\mathrm{voice}}$ from altering the Causal LLM representations, ensuring that LLM optimization remains strictly driven by speech generation gradients through the conditioning pathways. The weighted norm $\|\cdot\|_w^2$ separately averages errors across speaker identity and pooled acoustic feature slices, assigned weights of $1.0$ and $0.25$, respectively. Furthermore, to mitigate exposure bias between ground-truth targets $z^*$ during training and generated latents $z$ at inference, we apply a scheduled sampling strategy that progressively increases the probability of conditioning on generated voice latents across both pathways (without backpropagation through sampling). The total objective is $\mathcal{L} = \mathcal{L}_{\mathrm{speech}} + \mathcal{L}_{\mathrm{voice}}$, where $\mathcal{L}_{\mathrm{speech}}$ covers acoustic generation and sequence termination.

\textbf{Two-Stage Inference Workflow.} VoTSpeech executes inference via a two-stage prefill procedure with prefix key-value (KV) cache reuse. First, the Causal LLM processes the instruction prefix $[c, \langle v\rangle]$ to extract $h_c$, allowing Voice DiT to sample the voice latent $z$—serving as a brief intermediate step to establish the voice tone. Next, after prepending $e_v$, the LLM resumes prefill over the target text $x$ leveraging cached instruction KV states, followed by autoregressive speech latent generation. By reusing prefix KV states, this setup seamlessly integrates voice prediction into the generation pipeline as a lightweight pre-computation step.

\section{Experiments}
\label{sec:experiments}

\begin{table*}[t]
    \centering
    \caption{Comparison with existing systems and direct fine-tuning. Bold and underline mark the best and second-best means.}
    \label{tab:main_results}
    \begingroup
    \normalsize
    \renewcommand{\arraystretch}{1.15}
    \setlength{\tabcolsep}{3pt}
    \begin{tabular*}{\textwidth}{@{\hspace{6pt}\extracolsep{\fill}}lccccccc@{\hspace{6pt}}}
        \hline
        & \multicolumn{4}{c}{Objective Evaluation}
        & \multicolumn{3}{c}{Subjective Evaluation} \\
        \cline{2-5} \cline{6-8}
        Model & APS & DSD & RP & CER $\downarrow$ & NMOS & EMOS & AMOS \\
        \hline
        Ming-Omni-TTS-0.5B & 84.9 & 72.2 & 53.9 & \textbf{2.48} & $3.71 \pm 0.27$ & $3.71 \pm 0.24$ & $3.75 \pm 0.25$ \\
        Qwen3-TTS-12Hz-1.7B-VD & \textbf{87.1} & \underline{76.0} & 55.2 & 2.99 & $\underline{4.08} \pm 0.21$ & $4.10 \pm 0.19$ & $\underline{4.06} \pm 0.20$ \\
        MOSS-VoiceGenerator & 73.1 & 70.2 & 58.9 & 5.03 & $3.86 \pm 0.22$ & $3.72 \pm 0.18$ & $3.70 \pm 0.19$ \\
        VoxCPM2 & 84.7 & 71.8 & 56.8 & \underline{2.58} & $3.80 \pm 0.25$ & $3.82 \pm 0.18$ & $3.82 \pm 0.17$ \\
        \hline
        dots.tts-soar (fine-tuning) & \underline{85.3} & 75.0 & \underline{64.3} & 2.71 & $3.87 \pm 0.23$ & $\underline{4.12} \pm 0.18$ & $3.96 \pm 0.19$ \\
        VoTSpeech (Ours) & 84.8 & \textbf{78.5} & \textbf{66.9} & \underline{2.58} & $\mathbf{4.12} \pm 0.22$ & $\mathbf{4.30} \pm 0.18$ & $\mathbf{4.26} \pm 0.19$ \\
        \hline
    \end{tabular*}
    \endgroup
\end{table*}

\begin{table}[t]
    \centering
    \caption{Ablation of dual-path voice conditioning and target design. Bold marks the best result in each column.}
    \label{tab:ablation}
    \vspace{5pt}
    \begingroup
    \fontsize{9}{11}\selectfont
    \renewcommand{\arraystretch}{1.15}
    \setlength{\tabcolsep}{3pt}
    \begin{tabular*}{\columnwidth}{@{\hspace{3pt}\extracolsep{\fill}}lccc@{\hspace{3pt}}}
        \hline
        Variant & APS & DSD & RP \\
        \hline
        VoTSpeech (Full) & 84.8 & \textbf{78.5} & \textbf{66.9} \\
        w/o LLM-side Voice Conditioning & 83.8 & 76.7 & 63.9 \\
        w/o VAE Target Features & \textbf{85.1} & 77.7 & 63.6 \\
        \hline
    \end{tabular*}
    \endgroup
\end{table}

\subsection{Experimental Setup}
\label{ssec:experimental_setup}

\textbf{Baselines and Setup.}
We compare VoTSpeech with established instruction-guided TTS systems, including Ming-Omni-TTS-0.5B \cite{mingomni}, Qwen3-TTS-12Hz-1.7B-VD \cite{qwen3tts}, MOSS-VoiceGenerator \cite{moss}, and VoxCPM2 \cite{voxcpm2}. Built upon the same pre-trained dots.tts \cite{dotstts} backbone, we additionally construct a direct fine-tuning baseline, denoted as \textbf{dots.tts-soar (fine-tuning)}. This baseline treats instruction $c$ as ordinary textual context preceding synthesis text $x$ and directly optimizes instruction-to-speech generation without an explicit voice target, Voice DiT, or intermediate voice latent. Its LLM hidden states condition the Acoustic DiT through the original sequence pathway, while setting the separate speaker-conditioning vector to zero.

\textbf{Dataset and Implementation Details.}
Models are trained on the Chinese corpus presented in Sec.~\ref{ssec:data}, comprising 1,543.52 hours of unique audio recordings (each file counted once regardless of multiple instruction annotations). Training uses 8 $\times$ NVIDIA A100 GPUs for 20k steps ($\approx 17$ epochs). Initial learning rates are $10^{-5}$ for the backbone LLM and $10^{-4}$ for voice-design modules. The Voice DiT features a 4-layer Transformer with hidden dimension 512 and 8 attention heads. For scheduled sampling during synthesis training, the probability of conditioning on generated voice latents is 0 for the first 6k steps, increases linearly to 0.75 over the subsequent 3k steps, and remains fixed at 0.75 thereafter. Voice latent sampling uses 2 integration steps during training and 16 at inference.

\textbf{Evaluation Metrics.}
For instruction compliance on the InstructTTSEval Chinese (ZH) subset~\cite{instructttseval}, all systems—including baselines—are uniformly re-evaluated using a single Gemini 2.5 Pro~\cite{Gemini2P} judge setup with fixed prompts to ensure strict fairness. Accuracy (\%) is reported across Acoustic-Parameter Specification (APS), Descriptive-Style Directive (DSD), and Role-Play (RP). Speech intelligibility is measured by Character Error Rate (CER) using a pretrained Paraformer-ZH ASR model~\cite{Paraformer}. For subjective evaluation, 24 listeners rate 10 randomly selected samples per system on a 5-point scale. We report Mean Opinion Scores (MOS) for naturalness (NMOS), expressiveness (EMOS), and instruction adherence (AMOS), assessing perceived audio quality, delivery style fit, and instruction compliance, respectively. Subjective scores are reported with 95\% confidence intervals (CIs) in Table~\ref{tab:main_results}. Higher values indicate better performance for all metrics except CER.

\subsection{Experimental Results}
\label{ssec:experimental_results}

\textbf{Main Results.}
Table~\ref{tab:main_results} summarizes overall performance. VoTSpeech achieves leading DSD (78.5\%) and RP (66.9\%) accuracy while maintaining competitive APS (84.8\%) and low CER (2.58\%). Subjectively, it also secures top scores in naturalness (4.12), expressiveness (4.30), and instruction adherence (4.26). Compared to direct fine-tuning, VoTSpeech yields marked gains in high-level tasks like DSD (+3.5 percentage points) and RP (+2.6 percentage points). These quantitative improvements, along with subjective gains in instruction adherence (+0.30), demonstrate the clear advantage of explicitly predicting a voice latent over implicit instruction fine-tuning, while low CER confirms that speech intelligibility remains well-preserved.

\textbf{Ablation Study.}
Table~\ref{tab:ablation} examines \emph{where} the voice latent is injected and \emph{what} its target encodes via two variants:
(1) \textit{w/o LLM-side Voice Conditioning}, which removes $e_v$ while Acoustic DiT receives both $h_c$ and $z$ via $g_v$. Performance degrades across all tasks (notably a 3.0 percentage-point drop in RP), proving that Acoustic DiT conditioning alone is insufficient and LLM-side injection is vital for sequence-level prosody planning.
(2) \textit{w/o VAE Target Features}, which retains full dual-path interface (with residuals $P_h, G_h$) but trains target $z^*$ using only static speaker embeddings. Despite a marginal APS gain (+0.3 percentage points), DSD and RP decline by 0.8 and 3.3 percentage points, confirming that dynamic acoustic supervision beyond speaker identity is essential for expressive style and role-play control.
Overall, the full model achieves superior performance across complex scenarios through dual-path exposure and hybrid target design.

\textbf{Discussion.}
Comparing our main results against direct fine-tuning and ablations highlights two core strengths of VoTSpeech. First, incorporating dynamic VAE features provides a richer acoustic representation than static identity embeddings alone. Second, routing this predicted latent through dual-path conditioning equips both the LLM and Acoustic DiT with explicit voice guidance, drastically boosting instruction adherence and expressiveness. Nevertheless, our current voice latent design remains an utterance-level global descriptor. Because temporal pooling removes frame-level details, $z$ itself does not perform sequence modeling; fine-grained temporal dynamics—such as localized pitch contours and pacing—still rely on the sequence modeling capacity of the LLM backbone rather than the latent itself. Furthermore, without explicit disentanglement, pooled VAE features can occasionally encode environmental recording traits alongside target timbre. Developing voice latents with fine-grained temporal sequence modeling and explicit factor disentanglement represents an important direction for future work.



\section{Conclusion}
\label{sec:conclusion}
VoTSpeech decouples text-driven voice design from speech generation via an explicit continuous voice latent. By combining a hybrid target latent with dual-path conditioning, it seamlessly unifies LLM prosody planning and DiT acoustic rendering. Comprehensive evaluations show that this paradigm significantly improves instruction adherence and expressiveness while maintaining high speech quality, establishing explicit voice modeling as a viable pathway for controllable speech synthesis.

\section{Compliance with Ethical Standards}
The anonymous listening test was voluntary and followed the Declaration of Helsinki. Informed consent was obtained, no personally identifiable information was collected, and only aggregate ratings are reported.





\begingroup
\fontsize{9}{10}\selectfont
\bibliographystyle{IEEEbib}
\bibliography{refs}
\endgroup

\end{document}